\documentstyle[prl,aps,epsf,twocolumn]{revtex}
\begin{document}

\newcommand{\qp}{q_\perp}
\newcommand{\qpvec}{{\bf q}_\perp}
\newcommand{\rpvec}{{\bf r}_{\perp}}
\newcommand{\rp}{r_{\perp}}
\newcommand{\rvec}{{\bf r}}
\newcommand{\zhat}{\hat{\bf z}}
\newcommand{\rhat}{\hat{\bf r}}

\def\pmb#1{\setbox0=\hbox{#1}%
  \kern-.025em\copy0\kern-\wd0
  \kern.05em\copy0\kern-\wd0
  \kern-.025em\raise.0433em\box0 }
\def\grad{{\pmb{$\nabla$}}}

\draft

\title{Vortex Dynamics and Defects in Simulated Flux Flow}

\author{M.C. Faleski, M.C. Marchetti and A.A. Middleton}
\address{Physics Department, Syracuse University, Syracuse, NY 13244}

\date{\today}

\maketitle

\begin{abstract}
We present the results of molecular dynamic simulations of 
a two-dimensional vortex array driven by a uniform current through
random pinning centers at zero temperature. We identify two types
of flow of the driven array near the depinning threshold. For weak disorder 
the flux array contains few dislocation and moves via correlated 
displacements of patches of vortices in a {\it crinkle} motion. As
the disorder strength increases, we observe a crossover to a spatially
inhomogeneous regime of {\it plastic} flow, with a very defective vortex array 
and a channel-like structure of the flowing regions. 
The two regimes are characterized by qualitatively different spatial
distribution of vortex velocities. In the crinkle regime the distribution
of vortex velocities near threshold has a single maximum that shifts to
larger velocities as the driving force is increased. In the plastic regime
the distribution of vortex velocities near threshold has a clear bimodal
structure that persists upon time-averaging the individual velocities.
The bimodal structure of the velocity distribution reflects the
coexistence of pinned and flowing regions and is proposed as a
quantitative signature of plastic flow.
\end{abstract}
\pacs{PACS: 74.60 Ge, 74.60 -w, 62.20 Fe}


\narrowtext

\section{Introduction}

The problem of nonlinear collective transport through random media has
attracted much theoretical and experimental attention due to the 
interesting 
spatio-temporal phenomena that arise in a variety of physical systems
from the competition between interactions and 
disorder. In particular, the dynamics of driven elastic media
that are distorted by disorder, but cannot ``break'', 
has been studied extensively, both theoretically and numerically.
At zero temperature these systems exhibit a sharp depinning transition 
from a
pinned state below a critical driving force $F_c$ to a sliding state 
above $F_c$.
The transition can be described as a critical phenomenon in terms of scaling 
laws and critical exponents. 
The elastic medium model can be used to
describe the dynamics of weakly pinned Abrikosov flux 
lattices \cite{blatter}, fronts of wetting fluids invading porous media
\cite{rubio} and
charge density waves (CDW's) in anisotropic conductors \cite{gruner,fisher}
over a wide range of length scales.
It is, however, expected to eventually break down
(particularly in lower dimensionality) at short length scales
since it yields unphysical regions of unbounded strains
\cite{coppersmith}.
In addition, the elastic model is inadequate for many physical systems
with strong disorder that exhibit a spatially inhomogeneous plastic
response without long-wavelength elastic restoring forces. 
These include strongly pinned flux lattices \cite{shobo,jensen,shi}, 
invasion of nonwetting 
fluid in porous media \cite{rubio}, Wigner solids in 2DEG 
and fluid flow down a rough incline \cite{narayan}. 
In these systems the competition between drive and disorder generates
defects (dislocations, phase slips) in the medium that can qualitatively
change the dynamics \cite{koshelev,vinokur,balents}. 
Collective transport in the presence of topological defects
is still poorly understood.

Magnetic flux arrays in type-II superconductors are an ideal system
for investigating nonlinear collective transport since by changing the 
applied magnetic field one can tune the strength of the intervortex 
interaction and observe a crossover from a regime of weak pinning,
well described by collective flux creep theories, to a regime
of strong pinning with spatially inhomogeneous flow \cite {blatter}. 
Evidence for
this comes from early simulations of two-dimensional flux lattices
by Jensen et al. \cite{jensen} and by Shi 
and Berlinsky \cite{shi}. In addition, a variety of 
transport phenomena observed recently in superconductors has been 
attributed to
the inhomogeneous plastic response of the flux array, including
a nonmonotonic dependence of the critical current on temperature
just below the flux lattice melting point (peak effect), an unusual 
current and field dependence of $1/f$ broadband noise and fingerprint
phenomena \cite{shobo,ling,kwok,beasley}. 
On the other hand, the experiments only probe flux motion 
indirectly through transport measurements. For this reason their
interpretation
is difficult and still controversial.
Numerical work can therefore be very valuable to gain insight
into this complex problem and to serve as a guide for future
theoretical work.

In this paper we report on simulations of the dynamics
of a two-dimensional Abrikosov flux array driven by a uniform current
through random pinning centers at zero temperature.
The focus of our work, which distinguishes it from previous numerical
work on the same \cite{jensen,shi,vinokur,dominguez} or
closely related \cite{narayan,cha,huse} systems, 
is on identifying various types of flow and establishing a
connection between the type of flow or response (``elastic'' versus
``plastic'') and the presence of flux lattice defects
and the shape of the macroscopic response as embodied, for instance,
in the V-I characteristics. This will be very useful for the
interpretation of experiments. While most of the results presented here
are somewhat qualitative, our long-term objective is to carry out
simulations for realistic parameter values that will allow a detailed
comparison with experiments.
It is well known that
short-wavelength defects, such as dislocations, play
a more important role in two, rather than in three, dimensions \cite{gingras}.
For this reason many of our results will not
apply directly to three-dimensional flux arrays. On the other hand,
the study of two-dimensional systems is valuable both because of
intrinsic interest and because in many experimental situations the
flux arrays can effectively be modeled as two dimensional.
Thermal fluctuations are generally important in flux flow experiments
and purely dynamical phenomena associated with the current-induced depinning 
of the vortices cannot be dissociated from thermal-induced softening of the 
lattice. In this paper we specifically consider the flux array dynamics
at $T=0$ with the objective of disentangling these two types of effects.

Below we present the results of simulations of a driven flux array for
both a low density and a very high density of point pinning centers.
In both cases we identify two types of response or flow of the
flux array near the depinning threshold and a crossover from one type 
to another as the disorder strength is increased. 
For sufficiently weak disorder the flux array contains very few defects 
and moves via correlated
displacements of patches of vortices. The dynamics is similar to 
that observed by
Hu and Westervelt \cite{hu} in magnetic bubble arrays.  
Following these authors' suggestion, we refer to this type of response
as {\it crinkle} motion. For stronger disorder the response near threshold
is {\it plastic}, with vortices flowing
around pinned regions. The flux lattice is very defective and we observe
channels of liquid-like vortex array flowing around solid-like
pinned regions. The topology of the channels is not, however, fixed in time.
Channels open and close continuously as the flux array is driven 
over the impurities and
all vortices
participate in the motion at one time or another near threshold.
As the disorder is further increased the individual channels become 
longer lived and for very strong disorder we observe 
a filamentary structure with a fraction of vortices that never
move on the time scale of the simulation. To characterize the different 
regimes we have studied in detail the spatial 
distribution of vortex velocities. In the crinkle regime the distribution
of vortex velocity near threshold shows a single maximum corresponding
roughly to the average
velocity of the array, though at any time, some vortices are moving with
velocity significantly greater than the average value. 
The plastic
flow regime is characterized by bimodal velocity distributions near 
threshold, indicating that the velocity is spatially
inhomogeneous, with both pinned and flowing regions.
We discuss the correlation between these qualitative features 
of the velocity 
distribution and the shape of the macroscopic V-I characteristic and suggest
that the shape of the velocity distribution may be used for a crude 
classification of the type of response.

\section{The Model}

The specific model considered here is essentially the same
as that studied in earlier simulations by Jensen et al. \cite{jensen},
by Shi and Berlinsky \cite{shi} and,
more recently, by Koshelev and Vinokur \cite{vinokur}.
The two-dimensional pancake vortices are modeled as 
point-particles with finite-range
interaction and overdamped dynamics, driven through randomly placed pinning
centers by a uniform force ${\bf F}$ proportional to the external current.
The equations of motion for the two-dimensional vortex positions $\rvec_i$
are given by 
\begin{equation}
\label{eq:motion}
\gamma_1{d\rvec_i\over dt}=-\sum_{i\not= j}^{N_v}\grad_iV_v(\rvec_i-\rvec_j)-
\sum_{k=1}^{N_p}\grad_iV_p(\rvec_i-{\bf R}_k)+{\bf F},
\end{equation}
where $\{{\bf R}_k\}$ denote the random positions of the $N_p$ 
pinning centers and $N_v$ is the total number of vortices.  Here,
$\gamma_1$ is the friction coefficient of a single vortex, which will be 
incorporated in our unit of time.
The repulsive intervortex interaction has finite range $R_c$ and yields
a force $-\grad V_v(r)=f_ve^{-r/R_v}\big(1-r/R_c\big)\rhat$ 
on the $i$th
vortex, with $R_v\leq R_c$. 
The results presented  below have been obtained with  
$R_v=R_c$. 
The second term on the right hand side of Eq. (\ref{eq:motion})
is the attractive pinning force of range $R_p$, given by 
$-\grad V_p(r)=-f_p\big(1-r^2/R^2_p\big)^2\rvec / R_p$.
The range $R_v$ of the intervortex repulsion is of the order of
the superconductor penetration length, $\lambda$, while the range $R_p$ 
of the pinning potential is of the order
of the superconductor coherence length, $\xi$.
In the absence of pinning the flux array forms a stable
triangular lattice of lattice constant $a_0$.

One of the difficulties in carrying out a detailed numerical study
of the dynamical response of this model system is the large number
of parameters to be varied. In the following we have chosen the range $R_p$
of the pinning potential as our unit of length and the maximum 
intervortex force $f_v$ as our unit of force. 
In all cases the vortex lattice is rather dense, with $n_v R_v^2\sim 8-9$,
where $n_v=1/a_0^2$ is the areal density of vortices, and soft, 
with $c_{66}\sim 0.271-0.334$, where $c_{66}$
is the shear modulus of the clean vortex lattice, in the absence of disorder.
The pinning centers are modeled as point pins, in the sense that 
$a_0, R_v>>R_p$. We have considered sets of parameters corresponding to
two rather different density of pins: (i) a dense array of overlapping 
pins, with $N_p/N_v=133$ and $n_pR_p^2\approx 2.8$ with $n_p$ the areal 
density of pins, 
and (ii) a dilute
array of nonoverlapping 
pins, with $N_p/N_v=0.5$ and $n_pR_p^2\approx 0.046$.
The specific parameter values used are given below.
In both cases we have varied the strength of the disorder by
varying the maximum pinning force, $f_p$.

The mean motion of the flux array is described by the drift velocity
in the direction $\hat{\bf F}$ of the driving force, given by
\begin{equation}
\label{eq:drift}
v_d(F)=<{1\over N_v}\sum_{i=1}^{N_v}{\bf v}_i\cdot\hat{\bf F}>.
\end{equation}
The angular brackets denote the average over disorder.
In the numerical calculation we average over impurity realizations
by performing a time average since as time evolves the flux
array samples different impurity configurations. 
The mean velocity is proportional to the voltage $V$ from flux
motion, while the driving force $F$ is proportional to the
driving current $I$. Curves of  $v_d$ versus
$F$ correspond therefore to the V-I characteristics
of the material.

\section{Elastic Response}

Even in the absence of driving force the random pinning produces both
elastic and plastic (dislocations) deformations of the lattice.
If topological defects are explicitly forbidden in the model,
the distortion induced by disorder can be described within elasticity 
theory.

Treating the disorder as a perturbation, it has been shown
that for $d<4$ order persists only in region
of linear size $R_c$ that are pinned collectively \cite{larkin}.
The pinning length $R_c$
can be estimated by an Imry-Ma argument \cite {ma} by assuming that in
the presence of disorder the flux array deforms elastically to take
advantage of the pinning wells. The elastic energy cost associated 
with displacing a region of linear 
size $R$ by a distance $R_p$ is 
$\delta E_{el}(R)\sim c_{66}(R_p/R)^2 R^d$ in $d$ dimensions,
where $c_{66}$ is the shear modulus of the flux lattice.
The corresponding pinning energy gain is 
$\delta E_p(R)\sim \sqrt{n_v\Gamma}(R/R_p)^{(d/2)}$ where 
$\Gamma$ is the variance of the disorder potential arising from
uncorrelated point pins, with $\Gamma\approx n_p(f_pR_p^3)^2$. 
For dimension $d>4$ the elastic energy always exceeds the pinning energy at
large distances and the ordered state is stable for small disorder. 
For $d<4$ disorder dominates beyond the length
$R_c$ where the elastic strains
induced by disorder are of order one, or 
$\delta E_{el}(R_c)\sim\delta E_p(R_c)$. The elastic medium is broken up
in domains of size $R_c$, given by
\begin{equation}
\label{eq:rc} 
R_c= {R_p^3c_{66}\over\sqrt{n_v \Gamma}},
\end{equation}
in $d=2$.
Alternatively, the collective pinning length can be defined following Larkin
and Ovchinnikov \cite{larkin} as the length scale where
the mean square displacement $<[u(r)]^2>^{1/2}$
induced by disorder is of the order of the range of the pinning potential,
i.e., $<[u(R_c)]^2>\sim R_p^2$. 
The estimate described above  neglects logarithmic corrections
and assumes point pins of range  $R_p<<a_0$. 

The Larkin-Ovchinnikov 
collective pinning theory applies provided $R_c>>a_0$.
In addition, if topological defects can occur in the lattice,
the mean
distance between such defects must exceed $R_c$.
While our simulations have been carried out for parameter values where the
above inequalities generally do not apply, it is useful to briefly
summarize some of the properties of driven elastic
media for comparison.
The force needed to depin a region of linear size $R$ 
can be found by equating the energy gain due to the external force
$\sim FR_p R^2$ to the pinning energy $\delta E_p(R)$ of the region
and is given by 
\begin{equation}
\label{eq:threshold}
F(R)\sim {\sqrt{n_v\Gamma}\over R_p^2R}.
\end{equation}
In the weak pinning regime where Larkin domains of size $R_c$ are pinned
collectively by disorder, the threshold force needed to depin the medium
can be estimated as the force needed to depin a Larkin domain,
$F_{T}^{coll}=F(R_c)\sim n_v\Gamma/(c_{66}R_p^5)$.

When $R_c\sim a_0$ or $\sqrt{\Gamma}/R_p\sim c_{66}R_p^2$,
the collective pinning theory breaks down and vortices 
are pinned individually. In this strong pinning regime the threshold force 
can be estimated as $F_{T}^{strong}=F(R\sim a_0 )\sim n_v\sqrt{\Gamma}/R_p^2$.
The disorder-induced 
displacement of the lattice exceeds the range of the pinning potential and
the disorder can no longer be treated as a perturbation.
When this displacement becomes of order $a_0$, the Fourier 
components of the pinning potential with the periodicity of the 
underlying lattice become dominant and change qualitatively the
nature of the pinning \cite{natterman,giamarchi}. In this case 
one needs to go beyond the simple dimensional estimates just described, as
discussed recently by Giamarchi and Le Doussal \cite{giamarchi}.

The question of when and how, as a function of disorder strength,
topological defects proliferate has been addressed recently by Gingras 
and Huse \cite{gingras} for a ferromagnetic random field XY model.
Dislocations allow a region of linear size $R$ to better adjust to disorder,
yielding a gain in pinning energy. Gingras and Huse argue that
a bound for the length scale $R_d$ where dislocations proliferate can be
obtained by equating the elastic energy cost of a dislocation 
$\sim c_{66}a_0^2\ln R$ to this pinning energy gain. If the
pinning energy gain is estimated again as 
$\delta E_p(R)\sim \sqrt{n_v\Gamma}R/R_p$, we obtain 
$R_d\sim (a_0^2/R_p^2)R_c$. Notice, however, that since the disorder-induced
displacements of the lattice in the presence of dislocations exceed $R_p$,
the pinning energy no longer grows linearly with displacement and this 
estimate is at best a lower bound of the actual
pinning energy. For this reason all we can really infer from this argument
is that in the context of weak collective pinning $R_d>R_c$.
The focus of our paper is not on determining the length $R_d$, 
but rather on the dynamics
of the driven system and on the proliferation or healing of dislocations
as a result of the competition between disorder, drive and interactions.
We can estimate the force needed to depin and heal dislocations separated
by a length $L_d$ as $F_{d}\sim F(L_d)$, where $F(L)$ is given in 
Eq. \ref{eq:threshold}, with the result $F_{d}=n_v\Gamma/(c_{66}a_0^2R_p^3)$.
We remark that this dimensional estimate for $F_d$ is identical to the 
``crystallization'' force $F_t$ of Koshelev and Vinokur. 

We can then distinguish three regions as a function of the disorder 
strength $\Gamma$, as shown in the schematic ``phase'' diagram
of Fig. 1.  These regions may or may not be separated by actual 
phase transitions.  For $\Gamma<(c_{66}R_p^3)^2$, $L_d>L_c>a_0$ and 
the pinning is collective. The driven medium responds 
elastically and the relevant threshold force for depinning
is the force $F_{T}^{coll}\sim\Gamma$ needed to depin a Larkin domain.
For $(c_{66}R_p^3)^2<\Gamma<(c_{66}a_0^2R_p)^2$, 
the pinning is strong since $L_d>a_0>L_c$
and vortices are pinned individually. The threshold
force is estimated as the force $F_{T}^{strong}\sim\sqrt{\Gamma}$ 
needed to depin a single vortex.
In both these regions the force $F_{d}$ needed to depin and heal dislocations
present in the lattice is smaller than the threshold force.
When $\Gamma>(c_{66}a_0^2R_p)^2$, then $L_d<a_0$ and 
the force $F_{d}\sim\Gamma$ exceeds the threshold force for depinning.
In this region one may have a  scenario of the type proposed by Koshelev
and Vinokur \cite{vinokur}, with a pinned disordered solid for 
$F<F_{T}^{strong}$, a region of plastic flow for $F_{T}^{strong}<F<F_{d}$
and finally a flowing solid with no topological defects for $F>F_{d}$.

\begin{figure}[hbt]
\epsfxsize=3.5in\epsfysize=3.5in\epsfbox{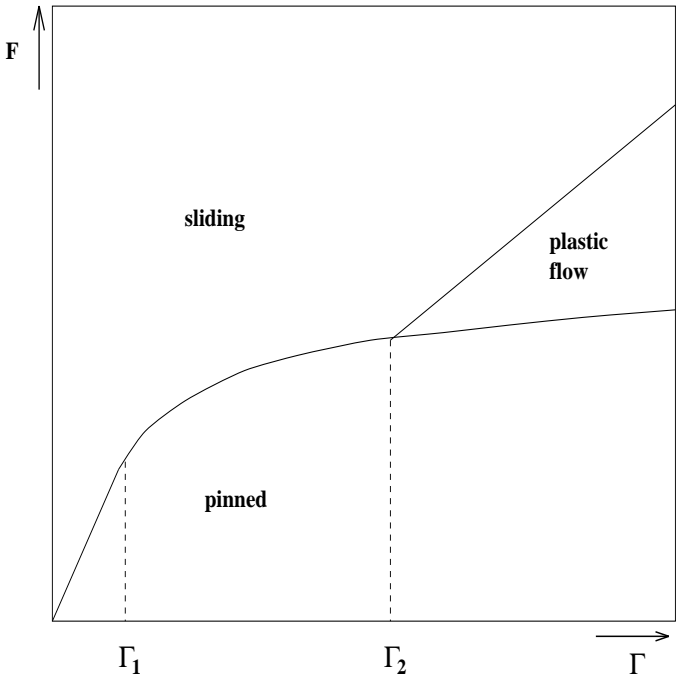}
\caption{A schematic ``phase diagram'' illustrating the various regimes
in the $(F-\Gamma)$ plane. The lines separating the various regions
represent the estimates of threshold force discussed in the text.
The boundary between pinned and sliding regimes is the collective
threshold force $F_{T}^{coll}\sim\Gamma$ for $\Gamma<\Gamma_1\sim (c_{66}R_p^3)^2$
and the strong pinning threshold force $F_{T}^{strong}\sim \Gamma^{{1 \over {2}}}$
for $\Gamma>\Gamma_1$. For 
$\Gamma>\Gamma_2\sim (c_{66}a_0^2R_p)^2$ there is a region of plastic
flow above the pinned region, separated from the sliding solid by
the force $F_{d}\sim\Gamma$.}
\end{figure}

The sliding state of an elastic medium driven through a random potential
can be studied analytically via a high-velocity perturbation theory.
The perturbation theory was introduced by Schmid 
and Hauger \cite{schmid} and by Larkin in the context of flux lattices
and further developed by Sneddon, Cross, and Fisher \cite{sneddon} for 
sliding CDW's.
More recently Zhu, Littlewood
and Millis discussed in detail the high-velocity perturbation theory 
for sliding Wigner crystals \cite{zhu}. 

The starting point of the perturbation theory is a description of the flux 
array
as an overdamped elastic continuum driven 
by an external force and distorted by short-range uncorrelated disorder.
The equation of motion for the two-dimensional displacement field
${\bf {u}}({\bf {r}},t)$ is given by 
\begin{equation}
\label{eq:elastic}
\gamma\partial_t {\bf {u}}({\bf {r}},t)=(c_{11}-c_{66})\grad(\grad\cdot{\bf 
{u}})
+c_{66}\grad^2{\bf {u}}+ {\bf {F}}_p({\bf{r}},t)+n_v{\bf{F}},
\end{equation}
where $\gamma$ is a friction per unit area, related to the single-vortex 
friction coefficient of Eq. \ref{eq:motion} by $\gamma=n_v\gamma_1$,
$c_{11}$ and $c_{66}$ are the compressional and shear elastic moduli 
of the two-dimensional flux lattice, and ${\bf F}_p$ is the pinning 
force per unit area,

\begin{equation}
\label{eq:pinforce}
{\bf F}_p({\bf {r}},t)=-\rho_0({\bf {r}})
\grad V({\bf {r}}+{\bf {u}}({\bf {r}},t)),
\end{equation}
where $\rho_0({\bf {r}})=\sum_{n=1}^{N_v}\delta({\bf {r}}-{\bf R}^0_n)$
is the spatially inhomogeneous density of the undistorted lattice, with
${\bf R}_n^0$ the sites of the triangular Abrikosov lattice.  The
coarse-grained quenched pinning potential, $V({\bf {r}})$ has
zero mean and short ranged correlations, 
$<V({\bf {r}})V({\bf r}')>=\Gamma({\bf {r}}) f({\bf {r}}-{\bf r}')$, with 
$f(r)$ 
a function that
drops rapidly to zero for $r>>R_p$.
For simplicity of notation we have neglected in Eq.
\ref{eq:pinforce} the nonlocality of the elastic constants. This can, 
however, be easily incorporated in the perturbation theory.
The drift velocity is defined here as
$v_d(F)=<\partial_t{\bf {u}}\cdot\hat{\bf F}>$, where the brackets
denote a spatial average, as well as a disorder average.
In the absence of disorder $v_d(F)=F/\gamma$. Treating the disorder
as a perturbation relative to the external driving force, one can then 
evaluate corrections to this uniformly sliding state. The details of the 
calculation are not given here as this follows closely the perturbation
theory for the Wigner crystal described recently by Zhu et al. \cite{zhu}.
Rather than expanding about the solution $v_d(F)=F/\gamma$ in the absence
of disorder, one actually constructs a self consistent perturbation
theory by writing ${\bf {u}}({\bf {r}},t)=v_dt+{\bf {s}}({\bf {r}},t)$,
solving for ${\bf {s}}({\bf {r}},t)$ in perturbation theory and then
requiring $<\partial_t{\bf {s}}({\bf {r}},t)>=0$. The lowest non-vanishing
correction $\delta v_d=v_d-F/\gamma$ to the mean velocity is given by,
\begin{eqnarray}
\label{eq:pert}
\delta v_d(F)\approx -{n_v^2 \over 2\gamma}& &\sum_{{\bf {G}}\not=0}
\Gamma(G)G^2
({\bf {G}}\cdot\hat{\bf F}) \nonumber\\
& &\times\int_{BZ}{d{\bf {k}}\over (2\pi)^2}
{\gamma{\bf {v}}\cdot{\bf {G}}\over (c_{66}k^2)^2+
(\gamma{\bf {v}}\cdot{\bf {G}})^2},
\end{eqnarray}
where ${\bf {G}}$ are the reciprocal lattice vectors of the triangular flux 
lattice and the ${\bf {k}}$-integral is over the first Brillouin
zone of size $k_{BZ}=\sqrt{4\pi n_v}$.
The disorder correlator is given by 
$\Gamma({\bf{G}})\approx \Gamma \approx n_p(f_pR_p^3)^2$, for $G<1/R_p$, 
and vanishes rapidly for $G>>1/R_p$. It therefore
cuts off the reciprocal lattice 
vectors sum at 
$G_{max}\sim 1/R_p$. For the case of point pins ($a_0>>R_p$),
the right hand side of Eq. \ref{eq:pert} can be evaluated exactly
by transforming the wavevector sum to an integral. 
It is, however,
more instructive to present the result in two limiting cases. 
If the velocity is not too large, $v_d<<c_{66}k_{BZ}^2R_p/\gamma$,
the main contribution to the ${\bf {k}}$ integral comes from the small $k$
region, corresponding to length scales much larger than the range of the
pinning potential. The upper limit of the wavevector integral can be extended 
to infinity, with the result,
\begin{eqnarray}
\label{eq:latticepert}
\delta v_d(F) &\approx& -{n_v\over 4\gamma c_{66}} \sum_{{\bf {G}}\not= 0}
\Gamma(G) G^2 {\bf {G}}\cdot\hat{\bf F}  {\rm {sgn}}({\bf {G}}\cdot{\bf {v}}_d)
\nonumber\\ &\approx& - {\Gamma n_v\over 4\gamma c_{66}R_p^5}.
\end{eqnarray}
In this intermediate velocity regime the lifetime of
elastic deformations
of the sliding medium is small compared to the time to cross the range of the
pinning potential, yielding collective pinning of the lattice.
For the two-dimensional case considered here in this regime one obtains
a force-independent correction to the drift velocity.
Conversely, if $v_d>>c_{66}k_{BZ}^2R_p/\gamma$, the time needed to cross
the range of the pinning potential - and therefore to see 
uncorrelated disorder -
is short compared to the lifetime of elastic deformations connecting 
neighboring vortices, and one obtains single-particle response, with
\begin{equation}
\label{eq:single}
\delta v_d(F)\approx -{\Gamma\over 16\pi\gamma_1 R_p^4}{1\over F}.
\end{equation}
Notice that the coefficient of $1/F$ in Eq. \ref{eq:single} is indeed
independent of vortex density.

\section{Numerical Results}

We have performed molecular dynamics simulations of arrays of 300, 920
and 1200
vortices using periodic boundary conditions. 
The results presented below are for two sets of parameter values, unless
otherwise specified. The data for the array with a dilute concentration
of pins ($N_p/N_v=0.5$) are obtained with $N_v=920$, $N_p=460$ and 
$R_v=9.9$. For these parameter values the clean Abrikosov lattice has
lattice constant $a_0=3.54$, with $n_vR_v^2=9.0$, and shear modulus 
$c_{66}=0.271$. As discussed earlier, all lengths are measured in units of 
$R_p$ and forces are in units of $f_v$.
The data for the densely pinned array ($N_p/N_v=133$) are obtained with 
$N_v=300$, $N_p=40,000$ and $R_v=20$. In this case $a_0=7.44$, with 
$n_vR_v^2=8.3$, and $c_{66}=0.334$. Table 1 shows the values of the 
collective pinning length $R_c$ given in Eq. \ref{eq:rc}
and the threshold force estimated using the dimensional analysis
discussed in section III for various pinning forces $f_p$.
For each set of parameters the value of the threshold force given 
in the table is the smaller of the two estimates $F_{T}^{coll}$ and $F_{T}^{strong}$.
\begin{center}
\centerline{Table 1}
\begin{tabular}{|ccc||ccc|} \hline
&$N_p / N_v =  0.5$ & & & $N_p / N_v = 133$ & \\ \hline \hline
$f_p$ & $R_c / a_0$ & $F_{T}^{est}$ &  $f_p$ & $R_c / a_0$ & $F_{T}^{est}$ \\ \hline
0.2 & 20.0 & $2.1$ x $10^{-4}$ &  0.03 & 7.8 & $1.4$ x $10^{-4}$ \\ \hline
0.5 & 13.3 & $1.3$ x $10^{-3}$ &  0.1 & 2.3 & $1.6$ x $10^{-3}$ \\ \hline
1.0 & 4.0 & $5.3$ x $10^{-3}$ &  0.3 & 0.8 & $1.0$ x $10^{-2}$ \\ \hline
1.5 & 2.7 & $1.2$ x $10^{-2}$ & 3.0 & 0.08 & $1.0$ x $10^{-1}$ \\ \hline
2.0 & 2.0 & $2.1$ x $10^{-2}$ & & &  \\ \hline
\end{tabular}
\vspace{12pt}
\end{center}

The drift velocity of the vortex array is shown as a function of driving 
force in Figs. 2a and 3a 
for various values of the maximum pinning force $f_p$. Figure 2 is for the 
sample with a low concentration of pins ($N_p/N_v=0.5$), while Fig. 3
is for the densely pinned sample ($N_p/N_v=133$). Figures 2b and 3b 
display the differential resistivity $dv_d/dF$.
Both Figs. 2a and 3a show a systematic evolution of the shape of the VI
curve with increasing disorder
strength not unlike that observed in the V-I curves of real superconductors 
\cite{shobo}.
For small pinning forces the velocity is nonlinear in $F$ only very near 
threshold, where it exhibits a very small region of negative
curvature. Correspondingly, there is no peak in the differential
resistivity. At larger pinning forces there is a change in the sign of
the curvature of the mean velocity that occurs at a value $F_{peak}$
above threshold, but
well in the nonlinear region, and yields a peak in the differential 
resistivity.
The location of this peak moves to larger driving forces as the pinning 
force increases. This dependence is particularly strong in the sample 
with a dense
pin array. 
For $F>F_{peak}$ the VI characteristic is concave down and as $F$ grows it
approaches the asymptotic value $v_d=F$, corresponding to a 
freely sliding array. In this region the
deviations from the linear behavior $v_d=F$ are fit {\it quantitatively}
by the single-particle perturbation theory result given in Eq. 
\ref{eq:single}. 
This can be rewritten as $v_d/F\approx 1-<F_p^2(0)>/F^2$, where
$<F_p^2(0)>=\Gamma/R_p^2$ is the mean square total pinning force.
For the specific pinning potential used in our simulations, 
$<F_p^2(0)>=(\pi/30) n_pR_p^2f_p^2$.
The rms velocity fluctuations defined as 
$v_{rms}=<[{1\over N}\sum_i{\bf v}_i\cdot\hat{\bf F}-v_d]^2>^{1/2}$
are also fit quantitatively by their single particle value,
$v_{rms}=[<F_p^2>/2N_v]^{1/2}$ in this region.
We stress that for very strong pinning the flux array can  be 
very disordered even in the region $F>F_{peak}$, with sometimes
as much as $50\%$ of the vortices with a coordination number
different from $6$ (see Fig. 4b below). This is because dislocations 
can be frozen in the sliding lattice in our $T=0$ simulations, yielding 
a disordered array that slides as a whole, with dislocations moving along 
at the same velocity as the rest of the lattice. This behavior may 
be a finite-size effect and is related to the hysteresis in
the defect configuration discussed below. 

\begin{figure}[hbt]
\epsfxsize=3.5in\epsfysize=3.5in\epsfbox{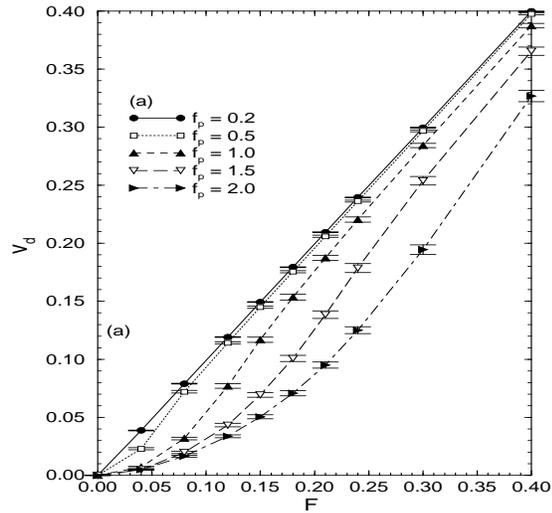}
\epsfxsize=3.5in\epsfysize=3.5in\epsfbox{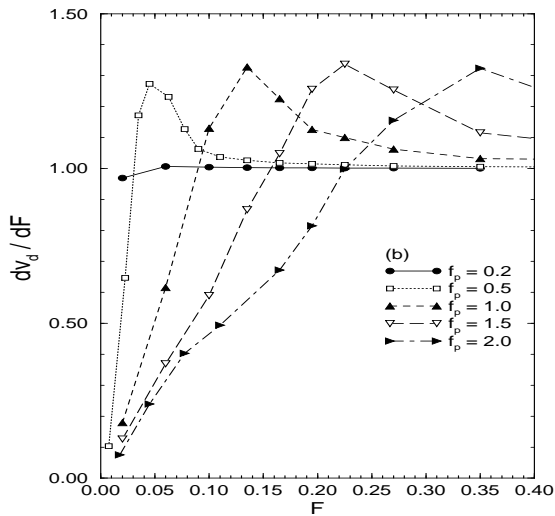}
\caption{Drift velocity (a) and differential resistivity $dv_d/dF$ (b)
versus driving force for the array with a dilute distribution
of pinning centers, $N_p/N_v=0.5$. The curves obtained by ramping the 
force up and down are indistinguishable. The error bars represent the 
value of $v_{rms}$.}
\end{figure}

\begin{figure}[hbt]
\epsfxsize=3.5in\epsfysize=3.5in\epsfbox{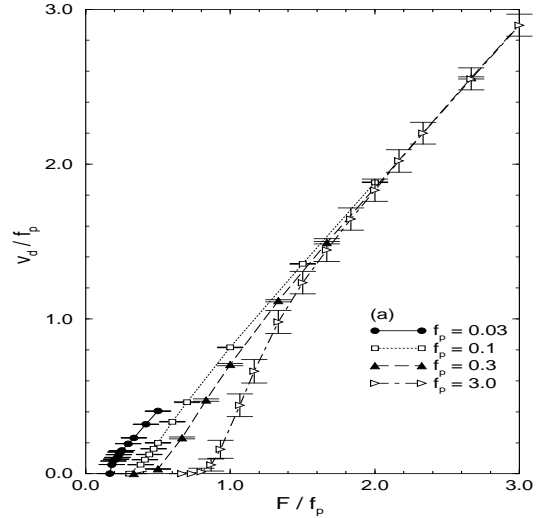}
\epsfxsize=3.5in\epsfysize=3.5in\epsfbox{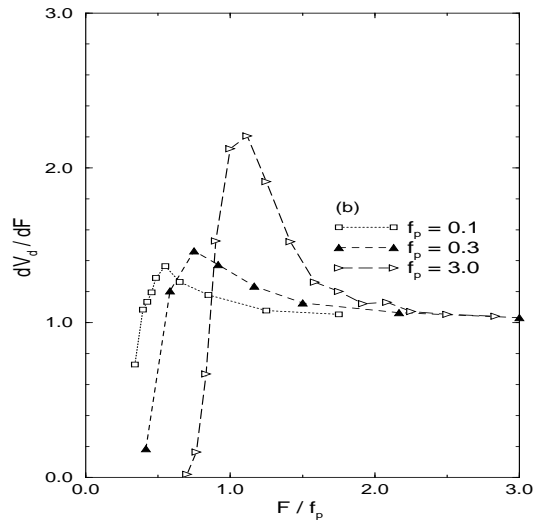}
\caption{Drift velocity (a) and differential resistivity $dv_d/dF$ (b)
versus driving force for the array with a dense distribution
of pinning centers, $N_p/N_v=133$. The curves obtained by ramping the 
force up and down are indistinguishable. The error bars represent the 
value of $v_{rms}$.  Notice that both $v_d$ and $F$ have been divided
by $f_p$ to display the data obtained for different values of the pinning 
force on the same graph.}
\end{figure}

For $N_p/N_v=0.5$ the threshold force remains very small
for all values of the pinning force studied. We cannot exclude that the
threshold force may vanish in this case, but we have not performed
extensive runs near threshold and finite-size scaling to determine
the location of the threshold precisely.
In fact, for a model with dislocations, it is not necessary that
$F_{T} \neq 0$ at zero temperature ($T=0$).
In contrast, the threshold force is large and clearly nonzero
for $N_p/N_v=133$. While we have not determined the threshold value 
accurately, we find that the numerical estimate agrees in order of
magnitude with the dimensional estimates given in Table 1.
Our numerical V-I characteristics resemble those obtained in
earlier simulations of the same model \cite{jensen,shi,vinokur}.
There is, however, an important difference between our results
and those of Koshelev and Vinokur \cite{vinokur}
in that we see no hysteresis in
the V-I characteristics other than that arising from finite size effects.
For small systems, we do see hysteresis in the V-I curves similar to
that reported by Koshelev and Vinokur, but this hysteresis vanishes
in larger systems.  This hysteresis may be due to the periodic boundary
conditions, which lead to metastable periodic attractors for the dynamics:
the hysteresis in small systems in our simulations is associated with such
periodic attractors.

We have also studied the evolution of the number of defects in the lattice
with driving force. In most of our runs the flux array is prepared
in an initial random disordered configuration and the driving force
is then ramped up from zero. We have tracked the defects in the lattice 
by doing  Voronoi constructions during the run and counting the number
of vortices that are not 6-fold coordinated. The 5- and
7-fold coordinated vortices are disclinations in the two-dimensional
triangular flux lattice and,
when paired, correspond to a dislocation. Fig. 4(a) shows the
time-averaged fraction
of 6-fold coordinated vortices as a function of driving force for
$N_p/N_v=0.5$. 
In looking at these curves it should be kept in mind that
the number of defects present at $F=0$ depends on initial conditions
and the realization of disorder.
As observed earlier \cite{vinokur}, we find that the flux array orders
at large driving force.
The value of driving force
where the number
of defects starts dropping is of the order of the location of 
the peak in the differential resistivity. Again,
we have observed no hysteresis in the number
of defects for the sample with a low concentration of pins
other than one arising from finite size effects. 
We do, however, find hysteresis in the number of defects when the disorder
is very strong, as shown in Fig. 4b. In this case when we ramp-up the force
from an initial disordered vortex configuration, defects get ``frozen in''
and the lattice never orders.
Starting from an ordered configuration at high fields, the lattice maintains
its order.
We have not been able to exclude that this hysteresis is
also a finite-size effect. We expect that this hysteresis 
will disappear in the presence of thermal fluctuations.
The value of driving force where the defects become frozen-in 
apparently corresponds
to the onset of the single-particle behavior discussed
below.

\begin{figure}[hbt]
\epsfxsize=3.5in\epsfysize=3.5in\epsfbox{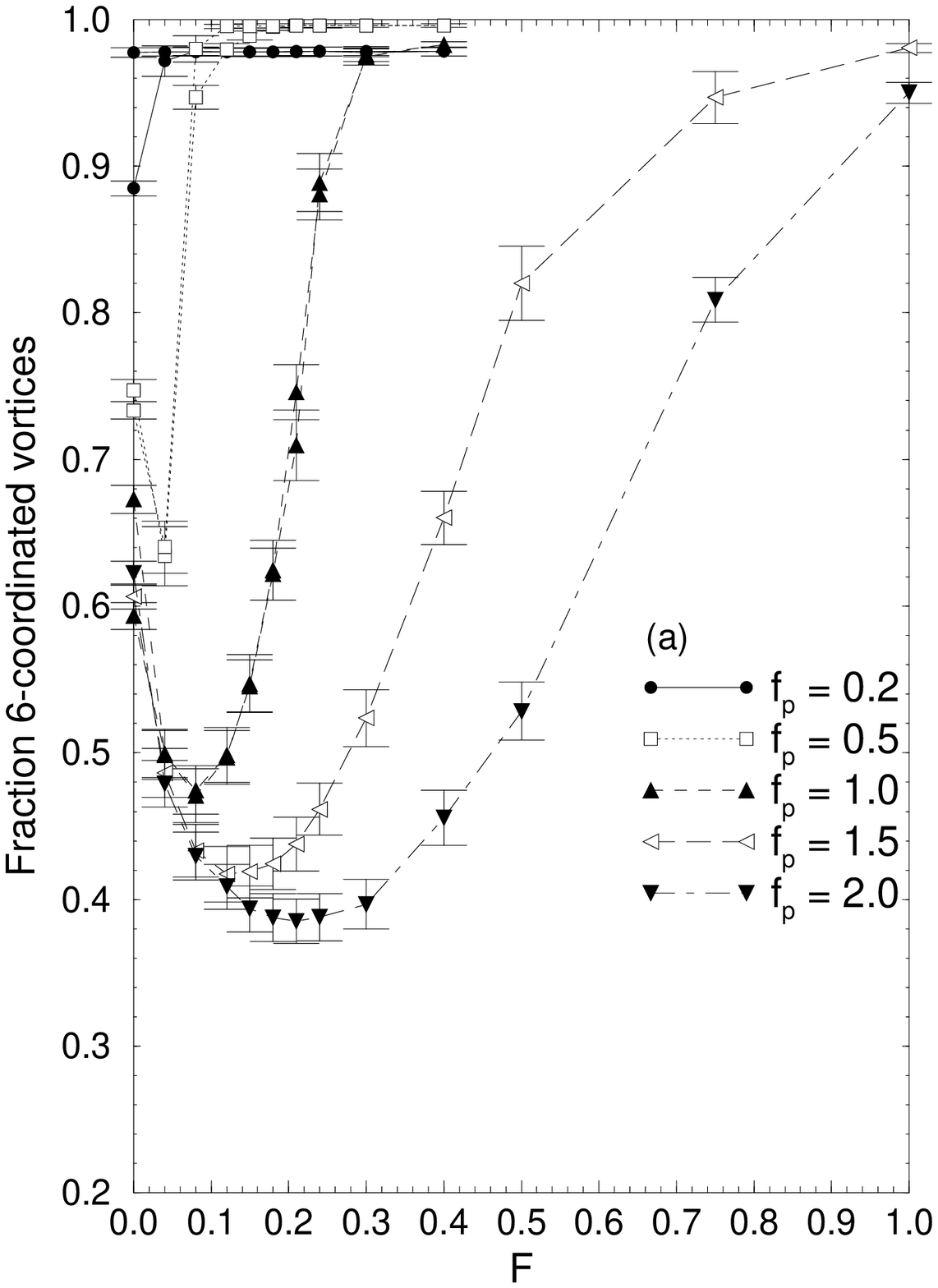}
\epsfxsize=3.5in\epsfysize=3.5in\epsfbox{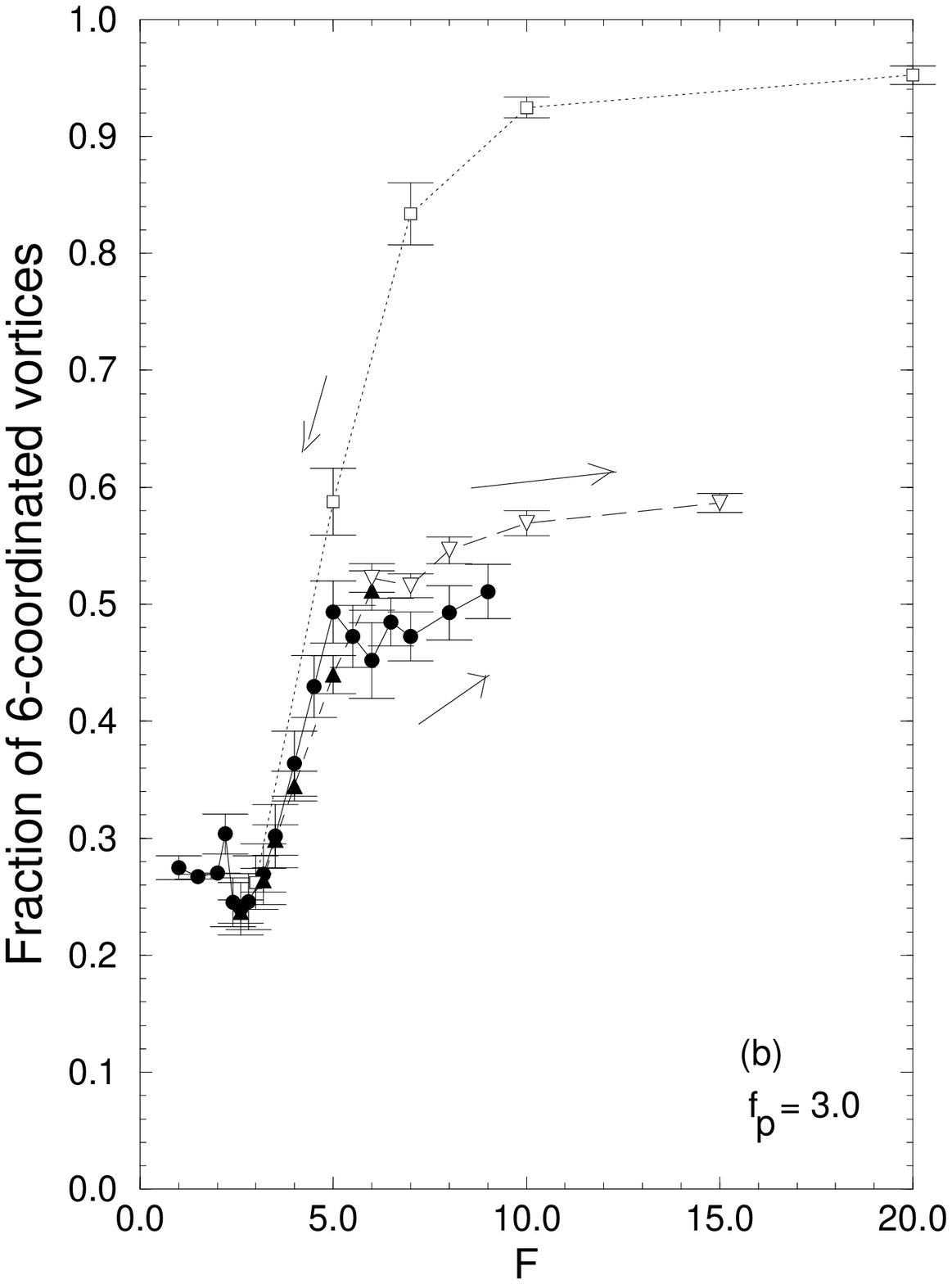}
\caption{Fraction of 6-fold coordinated vortices versus driving force.
Fig. (a) is for $N_p/N_v=0.5$ and the parameter values of Fig. 2.
In this case the curves obtained by ramping the force up and down 
are virtually indistinguishable and no hysteresis is observed.
Fig. (b) is for $N_p/N_v=133$, with the parameter values of Fig. 3
and $f_p=3$. The lower curves are obtained by ramping up the force from
an initial disordered configuration of the flux array. Data for both
$N_v=300$ (circles) and $N_v=1200$ (triangles) are shown to display 
the finite size effect. The upper curve (square) is obtained by ramping
down the force from an initial ordered configuration with $N_v=300$.}
\end{figure}

In order to correlate the macroscopic response of the vortex array
with the details of the microscopic vortex motion, we have performed
a variety of visualizations and we have studied the spatial distribution 
of vortex velocities.
One method of displaying the data that we have found useful is to plot
histograms of the component of the vortex velocity in the direction of the
driving force ($x$-component).
Figure 5 shows the evolution of such histograms 
of the $x$-component of the {\it instantaneous} vortex velocity near threshold
with driving force
for parameter values that yield crinkle flow. For all driving forces, 
the histograms
have a single maximum at a value of velocity close to the mean drift velocity.
The location of the maximum moves to larger velocities as the  driving
force increases.
Very close to threshold velocities are distributed asymmetrically about the
mean value and the histograms are not unlike those obtained from a 
phase-only model of CDW's \cite{S. coppersmith}. 
As the driving force increases the velocity distribution
becomes sharper and symmetric. 

\begin{figure}[hbt]
\epsfxsize=3.5in\epsfysize=3.5in\epsfbox{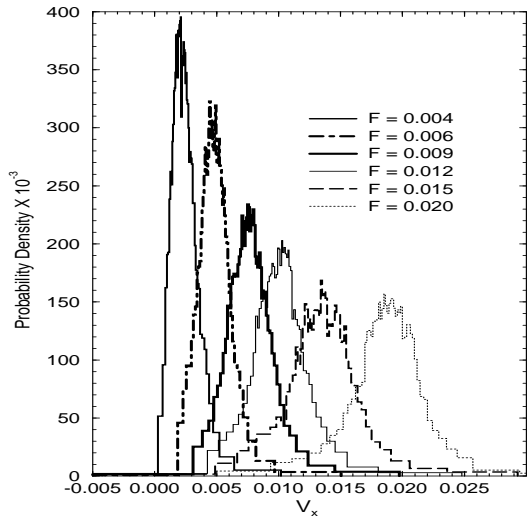}
\caption{Evolution of histograms of instantaneous velocity $v_x$ with applied
driving force for the parameters of Fig.2 and $f_p=0.2$.  The system exhibits
crinkle flow near threshold.}
\end{figure}

The instantaneous velocity distributions are quite different
for parameter values where
plastic flow is found. In this case the velocity histograms display a 
clear bimodal structure
near threshold, as shown in Fig. 6, indicating the presence of
two distinct ``typical'' velocities of the vortices.
The first peak, which is approximately centered at zero very near threshold,
is determined by the ``slow'' vortices in the array, located in pinned or
temporarily  pinned regions.
This peak has finite width due to the oscillations of the ``slow'' or
pinned vortices about the pins due to interactions with vortices flowing
nearby.
The peak at larger velocity is determined by
the ``fast'' vortices that flow in channels around the pinned regions.
We stress, however, that individual vortices are sometimes ``slow'', 
sometimes ``fast''.

\begin{figure}[hbt]
\epsfxsize=3.5in\epsfysize=3.5in\epsfbox{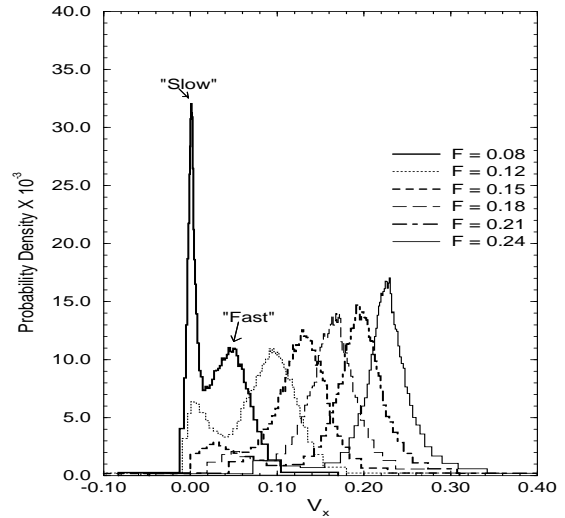}
\caption{Evolution of the histograms of instantaneous velocity $v_x$
with applied force for the parameter values of Fig. 2 and $f_p=1$. 
The system exhibits plastic flow near threshold and the histograms
clearly display a bimodal structure in this region.}
\end{figure}

As a result of the temporal fluctuations in the velocity, we expect
a large voltage noise in
this region, perhaps related to the experiments
by Marley, Bhattacharya and Higgins \cite{shobo}.
The various curves correspond to different driving forces, 
ranging from close to threshold (the threshold force for this case 
is estimated as $F_{T}\sim 5.3$x$10^{-3}$)  
to well within the linear regime.
The bimodal structure disappears at a value of $F$ close to the 
location $F_{peak}$ of the peak of the differential 
resistivity (here $F_{peak}\sim 0.125$). Beyond this value
the V-I curve rapidly becomes linear and the velocity distribution
is narrow and symmetric, centered at the mean velocity.
The origin of the maximum in the differential resistivity can easily
be traced back to the shape of the velocity histograms by studying
the location of the two peaks and their relative weights (Fig. 7)
as functions of the driving force. Using a crude approximation, we can write 
the drift velocity of the vortex array as $v_d=n_sv_s+n_fv_f$,
where $n_s$ and $n_f$ are the fraction of slow and fast vortices,
respectively, identified with the area under the respective peak of the
velocity distribution and shown in Fig. 7(b). Similarly, 
$v_s$ and $v_f$ are the corresponding velocities,
identified here with the location of the two peaks and displayed in Fig. 8a.
Using $n_s+n_f=1$, we obtain $(dv_d/dF)\approx (dn_f/dF)(v_f-v_s)$.
For $F<F_{peak}$ the slow vortices are essentially not moving ($v_s\approx 0$)
while the 
\begin{figure}[hbt]
\epsfxsize=3.5in\epsfysize=3.5in\epsfbox{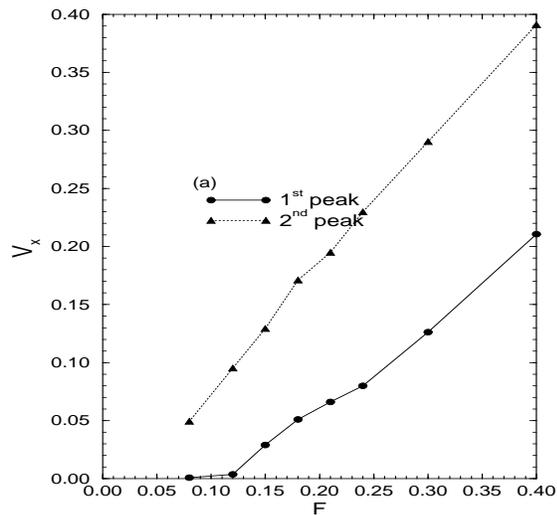}
\epsfxsize=3.5in\epsfysize=3.5in\epsfbox{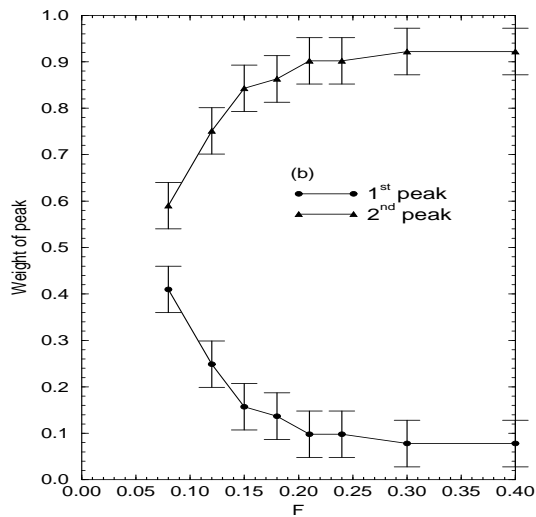}
\caption{These figures display the evolution with applied force of 
the location (a)  and weights (b) of the peaks of the histograms of Fig. 6.
The location is obtained by recording the velocities at which each peak is
maximum.  The weight of the $1^{st}$ peak is the ratio of the area under the 
part of the histogram up to the minimum between the two peaks,
to the total area under the curve.  The weight of the $2^{nd}$ peak is $1-$ 
the weight of the $1^{st}$ peak.}
\end{figure}

\noindent
number $n_f$ of fast vortices is increasing rapidly,
leading to a superlinear V-I. Above $F_{peak}$ the rate $dn_f/dF$
at which the slow vortex fraction decreases and the slow vortex fraction
increases slows down considerably,
while $(v_f-v_s)\sim F$ - it is this slowing down of the rate $dn_f/dF$
with increasing $F$ that is responsible for the peak in the differential 
resistivity.

The bimodal structure of the velocity histograms discussed above reflects
the spatial inhomogeneity of the instantaneous vortex velocity.
A crucial question for the characterization of plastic flow is whether
or not this bimodal structure will persist when the vortex velocity is
time averaged over times larger than those corresponding to an average
displacement of the lattice of at least a lattice constant.
It has been suggested that an important distinction between plastic and
elastic response can be found in the correlations of the time-averaged
velocity \cite{shobo}. 
Indeed in a model where dislocations are forbidden and
the response is therefore elastic, the time-averaged velocity will be
spatially homogeneous and correlated over the entire system size.
In contrast, in a system exhibiting plastic flow the time-averaged
local velocity should still be spatial inhomogeneous, yielding
bimodal structure of the corresponding histogram.
We have constructed histograms of 
time-averaged vortex velocities, defined as 
$\overline{v_i(T)}=\int_0^T{dt\over T}v_i(t)$ for various values
of $T$, where $T=1$ yields the histogram of instantaneous velocity
discussed above. The histograms are shown in Fig. 8 for two values of
the driving force. The bimodal structure clearly remains for the
time-averaged velocities.

To summarize, our model of a two-dimensional flux array driven
through quenched disorder exhibits two types of response.
For very weak disorder strength, the flux array exhibits ``crinkle''
motion, with correlated patches of vortices making small forward
jumps at different times, like a tablecloth being pulled
on a rough surface. At very small driving forces the lattice contains 
an appreciable number of
5- and 7-fold disclinations.
The defects  are concentrated at the boundaries between
the patches with correlated velocities and their number drops rapidly to zero
with increasing driving force. The distribution of
vortex velocities displays a single maximum at a value
of the velocity
of the order of the mean velocity of the array (Fig. 3).
As the driving force is increased and the flux array gets depinned,  the
maximum shifts to a higher velocity and the distribution broadens with
no substantial change in shape. The histograms of time-averaged velocities
become sharper as the averaging time increases and stop changing once the 
averaging time exceeds the time over which the vortex lattice advances
on the average a distance of a few $R_p$. 
This type of response is similar to that observed by
Hu and Westervelt in magnetic bubble arrays \cite{hu}. These authors report 
observing a bimodal velocity distributions, but this is because 
they only look at the distribution of velocity over a very small time

\begin{figure}[hbt]
\epsfxsize=3.5in\epsfysize=3.5in\epsfbox{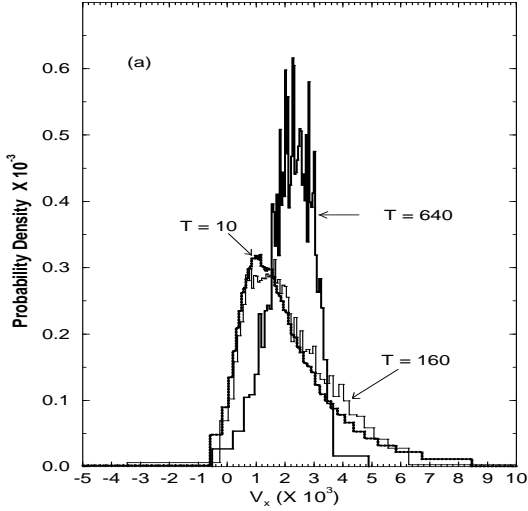}
\epsfxsize=3.5in\epsfysize=3.5in\epsfbox{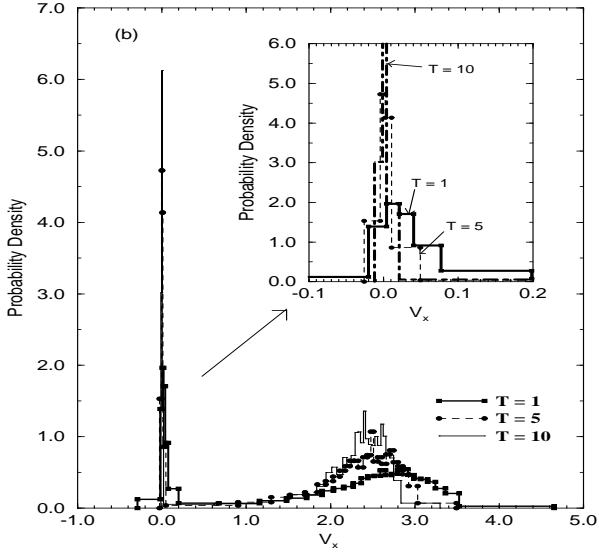}
\caption{These figures compare histograms of {\it instantaneous} and
{\it time-averaged} vortex velocities. The vortex velocities are averaged 
over a time $T$. The value $T=1$ yields the instantaneous velocity
histograms of Figs. 5 and 6. Figure (a)  corresponds to  parameter 
values yielding ``crinkle'' flow (same as Fig. 3, for $f_p=0.03$ 
and $F=0.0054$). In this case the instantaneous velocity histogram
exhibits a single maximum near $v_d\sim .001$ and is essentially 
identical to the histogram obtained with $T=10$ and $T=160$.
Notice that in a time $T=10$ the vortices displace on the average a
distance $\Delta x\sim 0.018 a_0$. The histogram for $T=640$
is sharper, but qualitatively unchanged, indicating that the system motion is
well correlated for these parameter values. The histograms shown in 
Fig. (b) refer to parameter values yielding plastic flow (same as Fig. 3
for $f_p=3$ and $F=3.5$). In this case the histogram of instantaneous 
velocities ($T=1$, solid curve) has a bimodal structure, which persists 
upon time averaging ($T=5$ or $\Delta x\sim 1.4 a_0$ and $T=10$ or $\Delta x 
\sim 2.7 a_0$), indicating true plastic response.}
\end{figure}

\noindent
scale, smaller than the time required for the array as a whole
to advance a distance of the order of the range of the pinning potential.
Hu and Westervelt also argue that the response exhibits scaling in this regime.
While we have not studied in sufficient detail the region near threshold
to observe scaling, it seems quite plausible that this ``crinkle''
regime will exhibit generic critical behavior analogue to that predicted 
and observed for an elastic medium, in spite of the presence of
defects. The fairly large number of defects present in our system at very 
small driving forces may very well be an artifact of our
initial conditions, with vortex positions chosen at random, rather
than equilibrated. It may be that if a low but finite temperature is
introduced in the model and the flux array is initially allowed to equilibrate
in the presence of the disorder, the number of defects present for 
the parameter values yielding crinkle motion would be practically
negligible even at the smallest driving forces.

In most of the region of parameter space studied we have observed plastic
flow of the flux array. This regime is characterized macroscopically
by a change in the sign of the curvature of the V-I characteristic well above 
threshold, which yields a maximum
in the differential resistance $dv_d/dF$, and by a large number of defects
in the region below the maximum.
The flow is spatially inhomogeneous. Over a short time interval 
one observes fluid-like flow of moving regions around pinned regions.
On the average, however, all vortices participate in the motion in 
the sliding state and no regions of the array are stuck for the entire
length of the simulation, even
near threshold. The evolution of the velocity distribution with 
driving force is shown in Fig. 4. There is clearly a bimodal velocity
distribution
in the sliding state which persists until all defects have healed 
and the V-I has become linear. A single vortex is in general ``slow'' 
for some of
the time and ``fast'' for some of the time. This should result in
a large voltage noise, as observed by Marley, Bhattacharya and Higgins
\cite{shobo}.

As the pinning force is increased, the persistence time of this
structure of pinned and flowing regions grows and pinned vortices
remain pinned for longer and longer times.
For the situation of strongest pinning among those studied ($f_p=3$ for 
$N_p/N_v=300$), we find that some vortices are pinned for the
entire length of the simulations, while others are flowing quite 
freely in channels surrounding the pinned regions. In this case the structure
of the channels is time independent near threshold. 
As the driving force is increased
all vortices are eventually depinned and the V-I becomes linear.
The array is very defective near threshold and the velocity distribution 
is bimodal and remains bimodal when velocities are time averaged.
The flux array displays a filamentary motion similar to that
observed by Middleton and Wingreen for an array of quantum dots
\cite{wingreen}, where near threshold the current flows in a single narrow 
channel and exhibits critical scaling.
Critical scaling was also predicted in a model of fluid flow down
a rough incline by Narayan and Fisher \cite{narayan}. In this case again
the flow pattern consists of directed channels that run across the system.
It may be that in this very strong pinning regime the driven flux array
also exhibits generic critical behavior, not unlike that of the fluid 
model of Ref. \cite{narayan}.

Some insight on the parameter regions where the two regimes described
above may be expected to occur can be gained by considering 
the Larkin-Ovchinnikov pinning length, given in Table 1 for the case where
the range $R_p$ of the pinning potential is small 
compared to both $R_v$ and $a_0$. 
If $R_v>>a_0$, the shear modulus of 
the flux array can be estimated as $c_{66}\sim f_v R_v/a_0^2$
and $L_c\sim R_v f_v\sqrt{N_v}/(f_p\sqrt{N_p})$.
The flux array should therefore be pinned collectively ($L_c>>a_0$)
provided $f_v\sqrt{N_v}/(f_p\sqrt{N_p})>1$. 
We find that these 
inequalities are generally satisfied in the cases
where we observe crinkle response.
Conversely, when $f_v\sqrt{N_v}/(f_p\sqrt{N_p})<<1$, even if  $R_v>>a_0$,
we obtain $L_c\sim a_0$ and vortices can be pinned individually
yielding plastic flow or eventually filamentary flow.
A transition between these two regimes is indeed observed as $f_p$ is
increased  for the parameter values discussed
above (here $L_c$ decreases over almost two orders of magnitudes over the range
of pinning forces studied, from $L_c\sim 20 a_0$ for $f_p=0.2$ to
$L_c\sim 0.3 a_0$ for $f_p=10.$). 
Using again the above estimate for $c_{66}$
for $R_v>>a_0$ we find that in the collective pinning regime
the threshold force needed to depin the array
is given by $F_{T}\sim (N_p/N_v)(R_p/R_v)(f_p^2/f_v)$. The increase
of the threshold force with $f_p$ observed in Fig. 2a is 
consistent with this dependence.

\vskip .2in
It is a pleasure to acknowledge helpful conversations with Shobo Bhattacharya.
This work was supported by the National Science Foundation 
at Syracuse through 
Grants DMR-9217284 and DMR-9419257, 
and at the ITP of the University of California in Santa Barbara 
through Grant PHY-8904035.  A. A. Middleton thanks
the Alfred P. Sloan foundation for support.
Finally, we thank Ki-Ho Lee for help with 
the numerical work in the early stages of this work.



\end{document}